\documentclass{bulsao}
\usepackage{graphicx}
\usepackage{latexsym}

\begin{document}

\title{Formation of Ionization-Cone Structures in Active Galactic Nuclei:
I. Stationary Model and Linear Stability Analysis}

\author{V.~L. Afanasiev$^1$, S.~N. Dodonov$^1$, S.~S. Khrapov$^2$, V.~V. Mustsevoi$^2$, A.~V.
Moiseev$^1$}

\institute{Special Astrophysical Observatory, RAS, Nizhnii Arkhyz,
Karachai-Cherkessian Republic, 357147 Russia \and Volgograd State University,
Volgograd, 400062 Russia}

\offprints{A.V. Moiseev, \email{moisav@sao.ru}}

\date{received:September 21, 2006/revised: November 24, 2006}

\titlerunning{Formation of Ionization-Cone Structures.. I.}

\authorrunning{Afanasiev  et al. }

\abstract{ We discuss  causes of the formation of the observed kinematics and
morphology of cones of ionized matter in the neighborhood of the nuclei of
Seyfert galaxies. The results of linear stability analysis of an optically
thin conic jet where radiation cooling and gravity play an important part
are reported. The allowance for radiation cooling is shown to result in
strong damping of all acoustic modes and to have insignificant effect on
unstable surface Kelvin--Helmholtz modes. In the case of
waveguide--resonance internal gravity modes radiative cooling suppresses
completely the instability of waves propagating away from the ejection
source and, vice versa, reduces substantially the growth time scale of
unstable sourceward propagating modes. The results obtained can be used to
study ionization cones in Seyfert galaxies with radio jets. In particular,
our analysis shows that surface Kelvin--Helmholtz modes and volume harmonics
are capable of producing regular features observed in optical emission-line
images of such galaxies. }

\maketitle

\section{Introduction}

Observations of many Seyfert galaxies in emission lines ($H_\alpha$, [OIII],
etc.) suggest that Narrow Line Regions  (NLRs) in these galaxies often have
conic shapes. The [OIII]/H$_\alpha$ line intensity ratio maps constructed by
some authors (see, e.g., \cite{pog}; \cite{fal2}) also suggest that gas in
NLRs is located inside cone-shaped (sometimes bipolar) structures with
opening angles $30^{\circ}- 110^{\circ}$ and linear sizes ranging from
several tens of parsecs to 20\,kpc (see Table~3 in the paper of \cite{wt}).

\begin{figure*}[!t]
\includegraphics[scale=1]{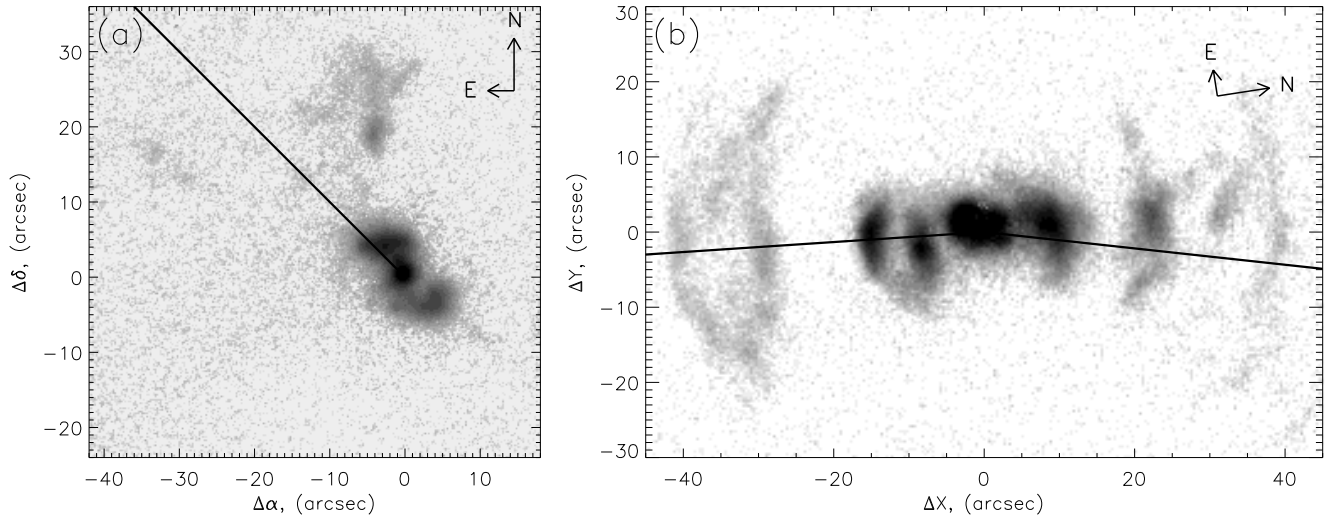}
\caption{ The [OIII]-line images of ionization cones with Z-shaped patterns
obtained from observations carried out at the 6-m telescope of the Special
Astrophysical Observatory of the Russian Academy of Sciences (\cite{my}). The
line indicates the orientation of radio jet according to Miyaji et al.
(1992)  and Wilson \& Tsvetanov (1994). (a)  NGC~3516, (b)
 NGC~5252. } \label{gal}
\end{figure*}

A number of authors thoroughly analyzed the structure of NLR in individual
galaxies, such as Mrk~3 (\cite{pogdr}; \cite{cap99}), Mrk~573 (\cite{tw};
\cite{fer}), NGC~3516 (\cite{miy}; \cite{veil}), NGC~5252 (\cite{wt}; Moorse
et al., 1998 ), and ESO~428-G14 (\cite{fal1}). The above authors point out
good correlation between the directions of ionized cones and those of radio
jets
--- their symmetry axes coincide to within $5-10^\circ$
(\cite{wt}; \cite{fal1}; \cite{nag}). Figure~\ref{gal} shows, as an example,
two [OIII] emission-line images of ionization cones in the galaxies NGC~3516
(Sy 1 type) and NGC~5252 (Sy 1.9 type) and the orientation of the axes of
their radio jets.

The orientation of the cones with respect to galaxy disks differs from one
galaxy to another, however, Wilson \& Tsvetanov (1994)  point out that in
late-type galaxies the cone symmetry axes are virtually perpendicular to the
disk plane, whereas in early-type galaxies they are inclined at small angles
to the disk.

The profile of narrow emission lines in the central regions of the
cone has a complex multicomponent structure. This is indicative of
several systems of gaseous clouds being observed at the
corresponding locations along the line of sight with mutual
velocities as high as several hundred  $km s^{-1}$
(\cite{cap99}; \cite{kais};  \cite{emsellem}).  %
In a number of cases, emission-line images of the galaxies considered
exhibit a Z(S)--shaped pattern (Fig. \ref{gal}) or other regular structures.
High spatial resolution HST images of Seyfert galaxies (\cite{cap96};
\cite{fal2}) indicate that the Z--shaped pattern begins in the circumnuclear
region (at galactocentric distances of 10--100~pc) and extends out to much
larger galactocentric distances (several kpc). The NGC~3516 galaxy
(Fig.~\ref{gal}a and \cite{miy}) shows Z--shaped filaments in the region
$r<10-14''$ ($\sim2$~kpc) with the Northeastern part of the cone extending
out to more than 6 kpc. The NLR in the NGC 5252 galaxy (Fig.~\ref{gal}b and
\cite{mor2}) also has a symmetric Z-like shape in the central region
$r<10-15''$ ($5-7$) kpc, whereas the biconical structure itself, which
contains regular emission-line ``arcs'' can be observed out to 18 kpc from
the center.

The motions of ionized gas inside cones are complex and poorly studied.
Two-dimensional velocity fields have been constructed in a few cases
(\cite{veil}; \cite{fer}; \cite{my}), which show that different portions of
Z-shaped patterns can be both ``blue-'' and ``redshifted'' with respect to
the nucleus. Moreover, in some galaxies, like, e.g., in NGC 5252, the
velocity fields in the region of ``arcs'' at large distances from the
nucleus exhibit emission-line filaments that are blueshifted exclusively in
one cone and redshifted exclusively in the diametrically opposite cone.

Ionization cones are believed to be due to the collimation of ionizing
radiation by the torus of matter accreting onto a supermassive black hole at
the nucleus of the galaxy. However, such a scenario of cone formation fails
to explain the presence of regular structures. Wilson (1993) argues that
ionized matter moves away from the galactic center, i.e., that it
constitutes a weakly collimated jet. Capetti et al. (1996) carried out a
detailed spectrophotometric study of such a Z-shaped emission-line region in
the Mrk~573 galaxy and pointed out that radiation of the nucleus is evidently
insufficient to produce the observed NLR and that an additional local
ionization source is required. Ferruit et al. (1999), who used panoramic
spectroscopy to analyze this object, also concluded that in Mrk~573 the
necessary additional contribution to ionization is provided by shocks
produced by the intrusion of the jet from the active nucleus into the
surrounding clouds of interstellar gas.

There is no consensus of opinion as to what causes the formation of regular
structures observed in ionization cones. The model of the interaction of the
jet with gaseous clouds in the circumnuclear region developed by Rossi et
al. (2000) explains a number of morphological features, but fails to describe
the development of symmetric Z-shaped features.

A number of researchers (see, e.g.,\cite{veil}; \cite{stef}) believe such
features to be helical shocks, which are due to the presence of a highly
collimated thin precessing jet. However, the hypothesis that this is the
case for all the objects discussed here appears to be too daring to say at
least. Mulchaey et al. (1992) interpreted the Z--shaped structure in NGC~3516
in terms of the bipolar outflow model where ejected gas is deflected toward
the galactic disk. Morse et al. (1998) explain the kinematic structure
observed in NGC 5252 by the presence of three ionized gas disks rotating in
differently tilted planes. In addition to the above scenarios there is the
possibility that excitation of a helical shock may be due to shear
instability, which develops at the layer between the matter of collimated
radio jet  and ionized gas moving at different velocities (\cite{fal1}).

We begin a series of papers where we naturally show with no additional
assumptions that the observed structures and velocity fields can be
explained in terms of the following scenario, which agrees with the unified
model of the activity of galactic nuclei (\cite{anton}; \cite{wil}):

\begin{itemize}
\item the highly collimated high-velocity bipolar jet (radio jet)
breaks through the torus of optically opaque matter accreting onto
the supermassive central object (black hole) in two diametrically
opposite directions parallel to the proper angular momentum of the
torus matter;

\item the jet matter squeezed by the ambient pressure heats up
intensively and expands rapidly toward the initial  ejection
through the narrow channel thus produced;

\item the internal  gravity waves propagating at an
angle to the jet axis undergo resonance superreflection at the
velocity shear surface made up of the jet boundaries;

\item  the harmonics of internal gravity waves resonate
between the jet boundaries and propagate inside the jet like in a
waveguide; during this process, the energy of gravity waves
increases with time, amplified via superreflection (resonance
waveguide instability);

\item the development of instability results in the formation of a
system of nonlinear waves around the jet, which heat up the
ambient medium; it is important that the wave resistance of the
ambient medium is significantly higher than that of the jet matter
and therefore the heating mentioned above occurs inside a cone with a
limited opening angle (which depends on particular parameters of
the system) around the jet;

\item the allowance for the possible nonlinear superposition of
different modes and projection effects permits obtaining the
qualitative pattern of the observed morphology of real objects and
of the velocity field inside the ionization cones.
\end{itemize}

As far as we know, no one has yet explored the possibility of the
development of the above modes, unlike the unstable acoustic modes of jets
emerging from young stellar objects, which were studied by a number of
authors (see \cite{ferrari}; \cite{pc}; Hardee \& Norman, 1988; \cite{ns};
Norman \& Hardee, 1988). In this paper, we show the principal possibility of
the buildup of both helical and pinch waveguide-resonance internal gravity
modes in conic jets, and discuss the results of nonlinear numerical modeling
of this process.

In Section \ref{2} we describe the equilibrium model employed; in Section
\ref{3} we give the linearized equations and formulate the problem of
determining the eigenfrequencies of unstable jet modes; in Section \ref{4}
we discuss the dispersion of perturbations during the linear stage of
instability, and in Section \ref{5} we summarize the main conclusions and
make the final comments. We analyze the results of nonlinear 2D- and
3D-modeling in our next paper (\cite{paper2}, hereafter referred to as
Paper~II).

\section{Equilibrium model}
\label{2}

Analyses of the dynamics of jet outflows from active galactic
nuclei naturally reveal three characteristic regions in radial
coordinate $r$:

\begin{itemize}
\item at  $r <(1-10)$ pc the gravitational field is determined
mostly by the Newtonian potential of the central massive object;

\item at $(1-10)\le r \le (500 - 1000)$ pc the jet is immersed in
dispersed mass of the stellar bulge, which can be considered to be
spheroidal to a first approximation. This region coincides with the
rigid-rotation region of the galactic disk (\cite{sofue})
 and therefore the gravitational potential can be assumed to
be proportional to squared radius to a fair accuracy;

\item  at $r > (0.5 - 1)$ kpc the variation of the gravitational
potential along the jet outflow differs substantially for
different galaxies. Moreover, it also strongly depends on the
orientation of the jet relative to the symmetry plane of the
galactic disk.
\end{itemize}

In this paper we analyze the spectrum of unstable modes of a jet located in
the gravitational field with potential  $\Psi \propto r^2$. Note that wave
structures that form in the jet in the inner part of the galaxy, where
potential can be fitted fairly well by the Newtonian formula, are incapable
of distorting the wave pattern over the jet region considered. First,
because the radial wavelength of perturbations in the field of a point mass
must decrease with  distance from the gravitating center, $\lambda_r \propto
r^{-1/2}$ (\cite{levin}), and the spatial scale length is incomparable with
the characteristic wavelength. Second, according to Levin et al. (1999), the
jet region, where energy can be fed to perturbations via resonance
superreflection and hence where wave structures with appreciable amplitudes
can exist, has a limited extent.

Our aim is to find out whether it is in principle possible for
unstable modes to develop at the layer between the jet and the
ambient medium, and therefore we do not incorporate the galactic
disk into our equilibrium model. We show in our next paper (Paper
II) that in the case of $\Psi(r) \propto r^2$  perturbations in
the ambient medium are localized in the conic domain near the jet.
Hence our formulation of the problem is formally correct, at
least for Seyfert galaxies with powerful bulges and ionization
cones lying outside the plane of the galactic disk.

We perform our analysis in the spherical coordinate system
$(r,\theta,\varphi)$, where the $\theta = 0$ axis coincides with
the symmetry axis of the jet with an opening angle of $\theta_j$
and outflow velocity of ${\bf V} = V_{j} {\bf e_r}$. Here ${\bf
e_r}$ is the unit basis vector. We model the medium by ideal gas
with the following equation of state
\begin{equation}
p_i = c_i^2 \rho_i / \gamma,
 \label{e1}
\end{equation}
where $p_i$ and $\rho_i$ are the unperturbed (equilibrium)
pressure and density, respectively; $c_i$ is the adiabatic sound
speed; subscript $i$ is equal to ``$j$'' and ``$a$'' inside and
outside the jet, respectively; we assume that adiabatic index
$\gamma$ is the same for the matter of the jet and the ambient
gas. We assume that gravitational field is spherically symmetric
with the center located at the coordinate origin. Variations of
the gravitational potential can be written in the following form:
\begin{equation}
\Psi = \Psi_0 + \frac{1}{2} \Omega^2 r^2, \label{e2}
\end{equation}
where $\Omega = const$ is the angular velocity of gas rotation in
the circumnuclear region of the disk and $\Psi_0 = const$ is a
normalizing constant.

We assume that gas outside the jet is at rest. We take into
account the possible heating of gas of the jet by the radiation
of the nucleus: $q_j
> 0$, where $q_j = \Gamma - \rho_j \Lambda$ is the amount of
energy absorbed by unit  mass per unit time; $\Gamma = \Gamma (T)$
and $\Lambda = \Lambda (T)$ are the heating and cooling functions,
respectively, which depend on temperature $T$ exclusively. In the
equilibrium state $q_a =0$ outside the jet.

Thus the spatial distribution of the model parameters that
characterize unperturbed flow has the following form:
\begin{equation}
V; \rho; c; q = \left\{ \matrix{ \quad 0 \, \,  ; \rho_{a}(r); c_{a}(r);
\quad 0 \, \, ,\quad \theta > \theta_j \cr V_{j}(r); \rho_{j}(r); c_{j}(r);
q_{j}(r),\quad \theta < \theta_j \cr }\right.
 \label{e3}
\end{equation}

We assume that the jet is contained by the pressure of ambient gas
and hence that the following equality is satisfied at $\theta =
\theta_j$:
\begin{equation}
\rho_{j}(r) c_{j}^2(r) = \rho_{a}(r) c_{a}^2(r).
 \label{e4}
\end{equation}

Note that below we make virtually no direct use of relation (\ref{e4}), its
fulfillment is required for realization of a flow with $V_\theta \equiv 0$.

Radial dependences in formula (3) are determined first and
foremost by the unperturbed balance of forces. Under the adopted
assumptions, the $r$ component of Euler's equations implies
\begin{equation}
{1 \over 2} {\partial V_i^2 \over \partial r} + \Omega^2 r = - {1 \over
\rho_i}{\partial p_i \over \partial r}.
 \label{e5}
\end{equation}

It follows from the continuity equation that
\begin{equation}
\rho_{j} V_{j} = \dot \mu_{j} / r^2. \label{e6}
\end{equation}

In equation (\ref{e6}) $\dot \mu_{j} = const$  is the mass-loss rate by the
system into a unit steradian. It is a free parameter of our model.

Finally, in the case considered the equation of energy balance, with
formulas (\ref{e1}) and (\ref{e6}) taken into account, can be written in the
following form:
\begin{equation}
V_i\ {\partial \over \partial r} \left( {V_i^2 \over 2} + {c_i^2 \over
{\gamma - 1}} \right) + \Omega^2 r V_i = q_i. \label{e7}
\end{equation}

Equation of state (\ref{e1}) closes equation set (\ref{e4})--(\ref{e7}).

We seek the solutions of this equation set in the power-law form: $f(r)
\propto r^{\alpha_f}$, where $f$ is any of the parameters that characterize
the system. It follows from equation (\ref{e5}), with equation (\ref{e1})
taken into account, that  $\alpha_V V_i^2 + \Omega^2 r^2 = - \alpha_p c_i^2
/ \gamma$. Hence we find:
\begin{equation}
\alpha_V = \alpha_c = 1,~ \alpha_\rho = - 3,~ \alpha_p =  - 1,~ \alpha_q = 2.
\label{e8}
\end{equation}

In this case, for a spherically symmetric potential the velocity
of matter in the jet relates to the sound speed in the ambient
medium as:
\begin{equation}
V_{j}^2 = {1 \over \gamma }\,(c_{j}^2 - c_{a}^2). \label{e9}
\end{equation}

The Mach number of the jet is $M = V_{j} / c_{j} < 1$, i.e., the
jet us subsonic. Note that this condition can be satisfied
simultaneously with $V_j / c_a \gg 1$, making it theoretically
possible for shocks due to the development of instability at the
jet boundary to exist in the medium that surrounds the jet,
because $c_{a} < c_{j}$.

Given that $\gamma > 1$, our model corresponds to an entropy
distribution $S_i$ that is stable against convective motions,
because
\begin{equation}
{d S_i \over dr} = {\rho_i^\gamma \over p_i}{d \over dr} \left({p_i \over
\rho_i^\gamma} \right) = {{\alpha_p - \gamma \alpha_\rho} \over r} = {{3
\gamma - 1} \over r} > 0. \label{e10}
\end{equation}

Substitution of power-law radial dependences into formulas (\ref{e5}) and
(\ref{e7}) and comparison with formula (\ref{e9}) yield:
\begin{equation}
V_j = {{\gamma (\gamma - 1)} \over {3 \gamma - 1}} {r q_j \over c_j^2}.
\label{e11}
\end{equation}

Hence the velocity of matter in the jet is unambiguously
determined by its temperature and heating due to external
radiation. Heating is rather important for outflows emerging from
active galactic nuclei, because the jet is illuminated
intensively by the radiation of the nucleus.

We finally determine, in view of formula (\ref{e8}), that the following
condition must be satisfied for the realization of the model constructed
above:
\begin{equation}
\Gamma (\varepsilon_i) = C_\Gamma \varepsilon_i,\qquad \Lambda
(\varepsilon_i) = C_\Lambda \varepsilon_i^{5/2}.
 \label{e12}
\end{equation}
Here $C_\Gamma$ and $C_\Lambda$ are constants and $\varepsilon_i$ is the
internal energy of gas. For further calculations, it is more convenient to
write heating and cooling in terms of internal energy, and not in terms of
temperature, by taking advantage of the fact that $(\gamma - 1) \varepsilon
= R T / \mu$ for ideal gas. Note that in the temperature interval $T<10^6$
the second relation in (\ref{e12}) agrees excellently with the dependence
$\Lambda (T) \propto T^{2.53}$ that is typical of ionized gas in Seyfert
galaxies (\cite{mac}; \cite{ns}). Linear temperature dependence of heating
function $\Gamma$ in formula (\ref{e12}) is also a good approximation.

\section{Linearized equations and formulation of the boundary-value problem}

\label{3}

Let us now analyze the stability of the model that we constructed
in the previous section against small perturbations. We proceed
from the following set of hydrodynamics equations :
\begin{equation}
{\partial{\bf V} \over \partial t} + ({\bf V \nabla}) {\bf V} = - {1 \over
\rho} {\bf \nabla} p - {\bf \nabla} \Psi, \label{e13}
\end{equation}
\begin{equation}
{\partial \rho \over \partial t} + ({\bf V \nabla}) \rho + \rho\,{\rm div}
{\bf V} = 0, \label{e14}
\end{equation}
\begin{equation}
{\partial \varepsilon \over \partial t} + ({\bf V \nabla}) \varepsilon +
(\gamma - 1) \varepsilon\,{\rm div} {\bf V} = C_\Gamma\,\varepsilon -
C_\Lambda\, \rho\,\varepsilon^{5/2}. \label{e15}
\end{equation}

We obtain the missing equation to close this set by choosing the
equation of state in the form $p = p(\rho,S)$ and computing its
derivative with respect to time:
\begin{equation}
\begin{array}{l}
{\displaystyle {d p \over d t} = \bigg({\partial p \over \partial
\rho}\bigg)_S {d \rho \over d t} + \bigg({\partial p \over \partial
S}\bigg)_\rho {d S \over d t} = }\\ \qquad {\displaystyle = c_i^2 {d \rho
\over dt} + {p_i \over c_v} {d S \over d t} = c_i^2 {d \rho \over d t} +
(\gamma - 1) \rho_i {T_i \over T} q. }
\end{array}
\label{e16}
\end{equation}

Our computations take into account the fact that $S = c_v \ln (p /
\rho^\gamma)$, $c_p - c_v = R / \mu$, $\gamma = c_p / c_v$, $d S /
d t = q / T$, and use the equation of state of ideal gas in the
form $p = R \rho T / \mu$.

Thus the last equation of the above set acquires the following
form:
\begin{equation}
\begin{array}{l}
{\displaystyle {\partial p \over \partial t} + ({\bf V \nabla}) p
= c_i^2 \bigg[ {\partial \rho \over \partial t} + ({\bf V \nabla})
\rho \bigg]  }\\    %\displaystyle
+(\gamma -1) \rho_i \varepsilon_i \left(C_\Gamma - C_\Lambda \rho
\varepsilon^{3/2}\right) .
\end{array}
\label{e17}
\end{equation}

To avoid doubts, note that subscript ``$i$'' indicates equilibrium
stationary parameter values.

We now use standard procedure of linear analysis to substitute pressure,
internal energy, density, and velocity of the medium in the form $f(r,
\theta, \varphi, t) = f_i(r) + \tilde f(r, \theta, \varphi, t)$, where
$\vert \tilde f \vert \ll f_i$. We assume that conditions
(\ref{e5})--(\ref{e7}) are satisfied to derive the following set of
equations describing the dynamics of small nonadiabatic perturbations
written for domains that are homogeneous in  $\theta$:
\begin{equation}
{\partial \tilde v_r \over \partial t} + V_i {\partial \tilde v_r \over
\partial r} + \tilde v_r {\partial V_i \over \partial r} = -
{1 \over \rho_i} {\partial \tilde p \over \partial r} + {\tilde \rho \over
\rho_i^2} {\partial p_i \over \partial r}, \label{e18}
\end{equation}

\begin{equation}
{\partial \tilde v_\theta \over \partial t} + V_i {\partial \tilde v_\theta
\over \partial r} + {V_i \tilde v_\theta \over r} = - {1 \over \rho_i r}
{\partial \tilde p \over \partial \theta}, \label{e19}
\end{equation}

\begin{equation}
{\partial \tilde v_\varphi \over \partial t} + V_i {\partial \tilde
v_\varphi \over \partial r} + {V_i \tilde v_\varphi \over r} = - {1 \over
\rho_i r \sin \theta} {\partial \tilde p \over \partial \varphi}, \label{e20}
\end{equation}

\begin{equation}
\begin{array}{l}
{\displaystyle
{\partial \tilde \rho \over \partial t} + {1 \over r^2} {\partial \over
\partial r} \left[ r^2 \left( \tilde \rho V_i + \rho_i \tilde v_r \right)
\right] }\\ {\displaystyle + {1 \over r \sin \theta}
\left[ {\partial \over \partial \theta} \left( \rho_i \tilde
v_\theta \sin \theta \right) + {\partial \over \partial \varphi}
\left( \rho_i \tilde v_\varphi \right) \right] = 0, }
\end{array}
\label{e21}
\end{equation}

\begin{equation}
\begin{array}{l}
{\displaystyle
{\partial \tilde p \over \partial t} + V_i {\partial \tilde p \over
\partial r} + \tilde v_r {\partial p_i \over \partial r} = c_i^2 \left(
{\partial \tilde \rho \over \partial t} + V_i {\partial \tilde \rho \over
\partial r}\right.
}\\ {\displaystyle +\left.\tilde v_r {\partial \rho_i \over
\partial r} \right) - p_i C_\Lambda \left(\varepsilon_i^{3/2} \tilde
\rho + {3 \over 2} \varepsilon_i^{1/2} \rho_i \tilde
\varepsilon\right), }
\end{array}
\label{e22}
\end{equation}

\begin{equation}
\begin{array}{l}
{\displaystyle {\partial \tilde \varepsilon \over \partial t} +
V_i {\partial \tilde \varepsilon \over \partial r} + \tilde v_r
{\partial \varepsilon_i \over \partial r} + \bigg[ (\gamma - 1)
{\partial V_i \over \partial r} - (\gamma + 1) {V_i \over r}  }\\
%\qquad
 {\displaystyle + {3 \over 2} C_\Lambda \rho_i
\varepsilon_i^{3/2} \bigg] \tilde \varepsilon + (\gamma -
1)\,\varepsilon_i\, \bigg[ {\partial \tilde v_r \over \partial r}
+ {2 \tilde v_r \over r}  }\\  %\qquad
{\displaystyle + {1 \over r
\sin \theta} \bigg( {\partial (\tilde v_\theta \sin \theta) \over
\partial \theta} + {\partial \tilde v_\varphi \over \partial
\varphi} \bigg) \bigg] = -\,C_\Lambda \varepsilon_i^{5/2} \tilde
\rho\,. }
\end{array}
\label{e23}
\end{equation}

We take equation (\ref{e11}) into account when deriving formula (\ref{e23})
from formula (\ref{e15}).

We seek the solutions for perturbed quantities in the following
form:
\begin{equation}
\tilde f(r,\theta,\varphi,t) = \hat f(\theta)\ r^{\beta_f}\, \exp{\{i \chi
(r, t) + i m \varphi \}}. \label{e24}
\end{equation}

Based on the requirement of conservation of the flux of perturbation energy
through a sphere of arbitrary radius  ($r^2 \tilde p \tilde v_r = const$),
we find the exponent of the radial dependences of perturbation amplitudes to
be $\beta_f = \alpha_f - \alpha_\Psi / 2$. Here $\alpha_f$ is, as above, the
exponent for equilibrium quantities, and $\alpha_\Psi = 2$, the exponent for
the gravitational potential (\ref{e2}). It is important that for this
$\alpha_\Psi$ value formula (\ref{e24}) gives exact solutions. Let us now
introduce the following designations: $k =
\partial \chi /
\partial r$, $\omega = - \partial \chi / \partial t$. In this case, equation set
(\ref{e18})--(\ref{e23}) for the equilibrium model considered reduces to the
following two ordinary differential equations:
\begin{equation}
{d \hat p \over d \theta} = \hat \omega_i (\hat \omega_i + \delta_i) \rho_i
r^3 \hat \xi, \label{e25}
\end{equation}
\begin{equation}
{d \hat \xi \over d \theta} = { 1 \over r \hat \omega_i (\hat \omega_i +
\delta_i)} \bigg[\lambda_i + {m^2 \over r^2 \sin^2 \theta}  \bigg] {\hat p
\over \rho_i} - \hat \xi \mbox{cot} \theta. \label{e26}
\end{equation}
Here $\hat \omega_i = \omega - k V_i$ is the Doppler shifted
perturbation frequency and $\delta_i = i V_i / r$, $\hat \xi$ is
the complex amplitude of perturbed Lagrangian
$\theta$--displacement of the medium such that:
\begin{equation}
\tilde v_\theta = {d \tilde \xi \over d t} = - i (\omega - k V_i) \tilde \xi
= - i \hat \omega_i \tilde \xi. \label{e27}
\end{equation}
We introduced the following designation in formula (\ref{e26}):
\begin{equation}
\begin{array}{l}
{\displaystyle
 \lambda_i = \left(k + {i \over r}\right) \left(k + {2 i \over r}\right)
 }\\ %\qquad
{\displaystyle
 -\bigg [ \rho_i (\hat \omega_i^2 - \delta_i^2) + {i
\over r} (k + {i \over r}) p_i \bigg ]
 }\\ %\qquad
{\displaystyle
 { \times \bigg [ {{\hat \omega_i - 2 \delta_i \over p_i} +
 {i \hat A_\Lambda \over r \rho_i}
 {(3 \gamma - 1) (k + {2 i \over r}) \over \hat \omega_i + \delta_i}}
 \bigg ]
}}\\ %\qquad
{\displaystyle
\times \bigg \{ {\hat \omega_i - (3 \gamma + 1) \delta_i }
 + \hat A_\Lambda \bigg [ (\gamma - 1) (\hat \omega_i - \delta_i)
}\\ %\qquad \qquad
{\displaystyle -
 i C_\Lambda \rho_i \varepsilon_i^{3/2} -
{3 \gamma - 1 \over \gamma r^2}
 {c_i^2 \over \hat \omega_i + \delta_i} \bigg ] }\bigg \}^{- 1},
%}
\end{array}
\label{e28}
\end{equation}
where
$$
\hat A_\Lambda = 1 - {{3 \over 2} i C_\Lambda \rho_i \varepsilon_i^{3/2}
\over \hat \omega_i - \delta_i + {3 \over 2} i C_\Lambda \rho_i
\varepsilon_i^{3/2}}.
$$

To solve the boundary-value problem for the determination of unstable-mode
eigenfrequencies, four boundary conditions must be satisfied. We use
standard boundary conditions at the jet axis based on the physical
assumptions of Hardee \& Norman (1988) and Norman \& Hardee (1988). All
displacements for axisymmetric perturbations should be equal to zero at the
jet axis, $\hat \xi (0) = 0$. In the case of nonaxisymmetric perturbations
($m
> 0$) the term with $\hat p (0)$ in formula (\ref{e26}) is finite
along the jet axis only for $\hat p (0)\equiv0$. Hence:
\begin{equation}
\left\{ \matrix{ \hat p (0) = 0,\ \ \ \ \ m \ge 1,\cr \hat \xi (0) = 0,\ \ \
\ \ m = 0,\cr}\right. \label{e29}
\end{equation}

\begin{equation}
\left\{ \matrix{ \hat p (\pi) = 0,\ \ \ \ \ m \ge 1,\cr \hat \xi (\pi) = 0,\
\ \ \ \ m = 0,\cr}\right.
 \label{e30}
\end{equation}

To find the boundary conditions at the external ($\theta=\theta_j$) jet
boundary, we integrate equations (\ref{e25}) and (\ref{e26}) from
$\theta_j-\varepsilon$ to $\theta_j+\varepsilon$ (where
$\varepsilon\rightarrow0$). Here we adopt
$\rho_i(\theta)=\rho_j-(\rho_j-\rho_a)\hat\theta(\theta-\theta_j)$, where
$\hat\theta(\theta-\theta_j)$ is the symmetric Heaviside step function. In
the same way we determine $V_i(\theta),\,p_i(\theta),
\varepsilon_i(\theta)$, and $c_i(\theta)$. Integration yields the following
conditions:
\begin{equation}
\hat p (\theta_j - 0) = \hat p (\theta_j + 0), \label{e31}
\end{equation}

\begin{equation}
\hat \xi (\theta_j - 0) = \hat \xi (\theta_j + 0),
 \label{e32}
\end{equation}
because the right-hand sides in equations (\ref{e25}) and (\ref{e26})
contain no $\delta$--functions (i.e., derivatives of $\theta$--functions).
From the physical viewpoint this means the continuity of total pressure
$(p_i+\tilde p)$ at the jet boundary bended by perturbations (with the
allowance for equation (\ref{e4})) and the absence of cross-boundary gas
flows.

The dispersion properties of small perturbations in the system
considered are fully characterized by the following dimensionless
parameters:
\begin{itemize}

\item constant Mach number $M = V_{j} / c_{j}$ along the jet;

\item density difference between the ambient gas and jet matter
$\tilde{R} = \rho_{a} / \rho_{j} = c_{j}^2 / c_{a}^2$;

\item radius-independent fractional radial wavenumber $kr$;

\item  helical-mode number (the number of spiral arms in azimuth)
$m$;

\item the ratio of the acoustic wave period to the relaxation time
scale of the gas of the jet $\tau = C_\Lambda\rho_j
\varepsilon_j^{3/2} / k c_j$;

\item dimensionless phase velocity of perturbations along the jet
axis $z = \omega / k c_{j}$, which is a solution of equation set
(\ref{e25})--(\ref{e26}) with boundary conditions (\ref{e29})--(\ref{e32}).
In this case, $z$ is also radius independent.
\end{itemize}

The condition of growth of fractional  perturbation amplitude
($Im\, z > 0$) means that the mode considered is unstable.

Only one of the two parameters --- $M$ and $\tilde{R}$ --- is independent,
whereas the second parameter can be computed by relation (\ref{e9}): $M^2 =
(1 - 1 / \tilde{R} ) / \gamma$. We further assume, for the sake of
simplicity, that gas has the same composition throughout the entire system,
implying that the external cooling coefficient $\tau_a$ is related to the
cooling coefficient in the jet as \mbox{$\tau = \sqrt{\tilde{R} }\, \tau_a$.}

Note that in the case of fixed medium with no radiative cooling ($V_i = 0$,
$C_\Lambda = 0$, $\delta_i = 0$, $\hat \omega_i = \omega$) we set $\tilde
v_{\theta} \equiv 0$, $m=0$ to immediately derive from the linearized
equation set for perturbations with wave vector ${\bf k}  \parallel {\bf
e_r}$  the following dispersion equation, which is equivalent to $\lambda_i
= 0$ in equation (\ref{e26}). Its solution has the form:
\begin{equation}
\omega = \pm\ k c_i \sqrt{1 + {4 \gamma - 1 \over \gamma k^2 r^2}} .
\label{e33}
\end{equation}
Thus in the short-wavelength approximation ($kr\gg 1$) we obtain
the common dispersion law for acoustic waves: $\omega \simeq \pm k
c_i$. The opposite limiting case $kr \ll 1$ leads to the following
dispersion relation:
\begin{equation}
\omega \simeq \pm \sqrt{4 \gamma - 1 \over \gamma} {c_i \over r}. \label{e34}
\end{equation}

Relation (\ref{e34}) describes the dispersion law for long-wavelength
gravity waves.

\begin{figure*}
\includegraphics[width=16 cm]{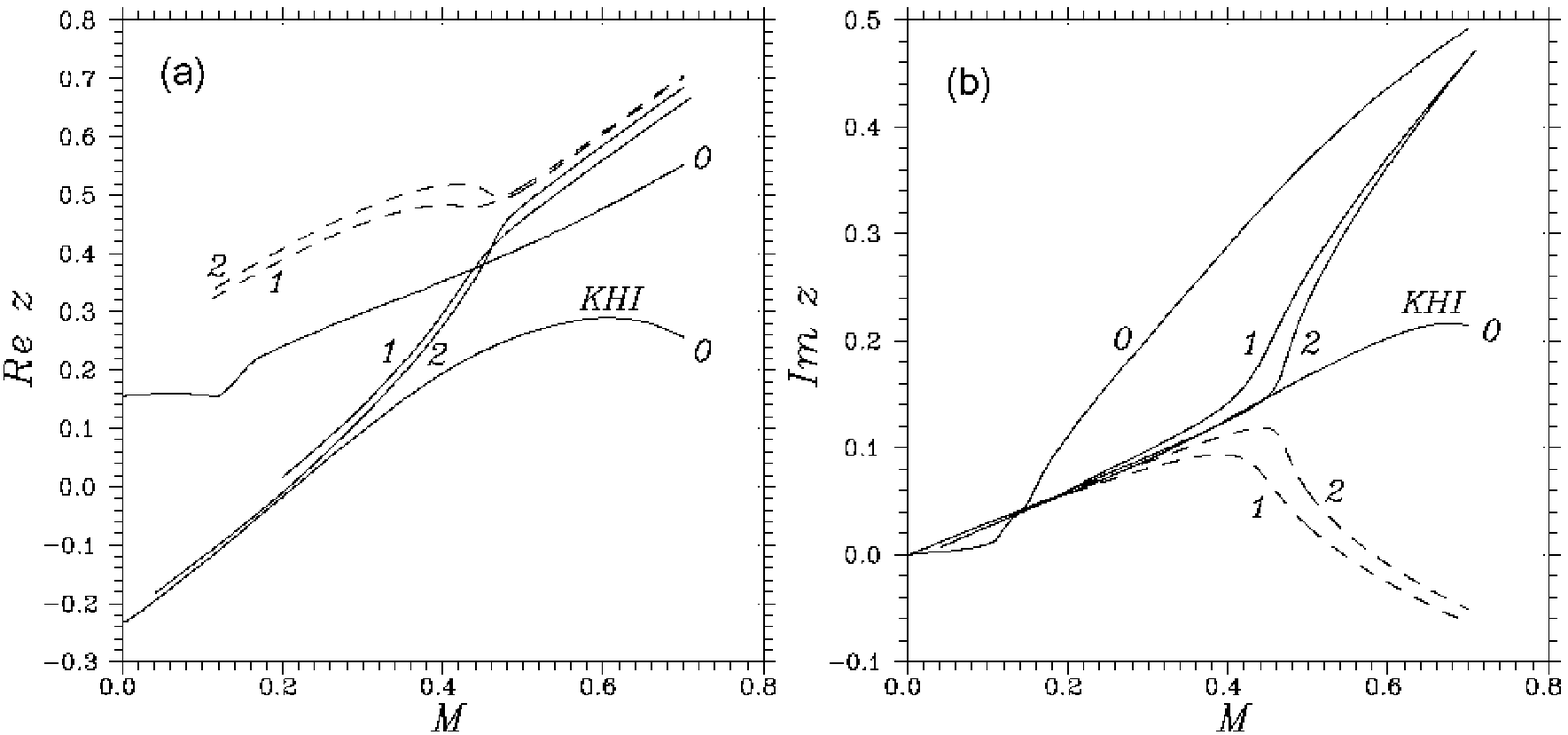}
\includegraphics[width=16 cm]{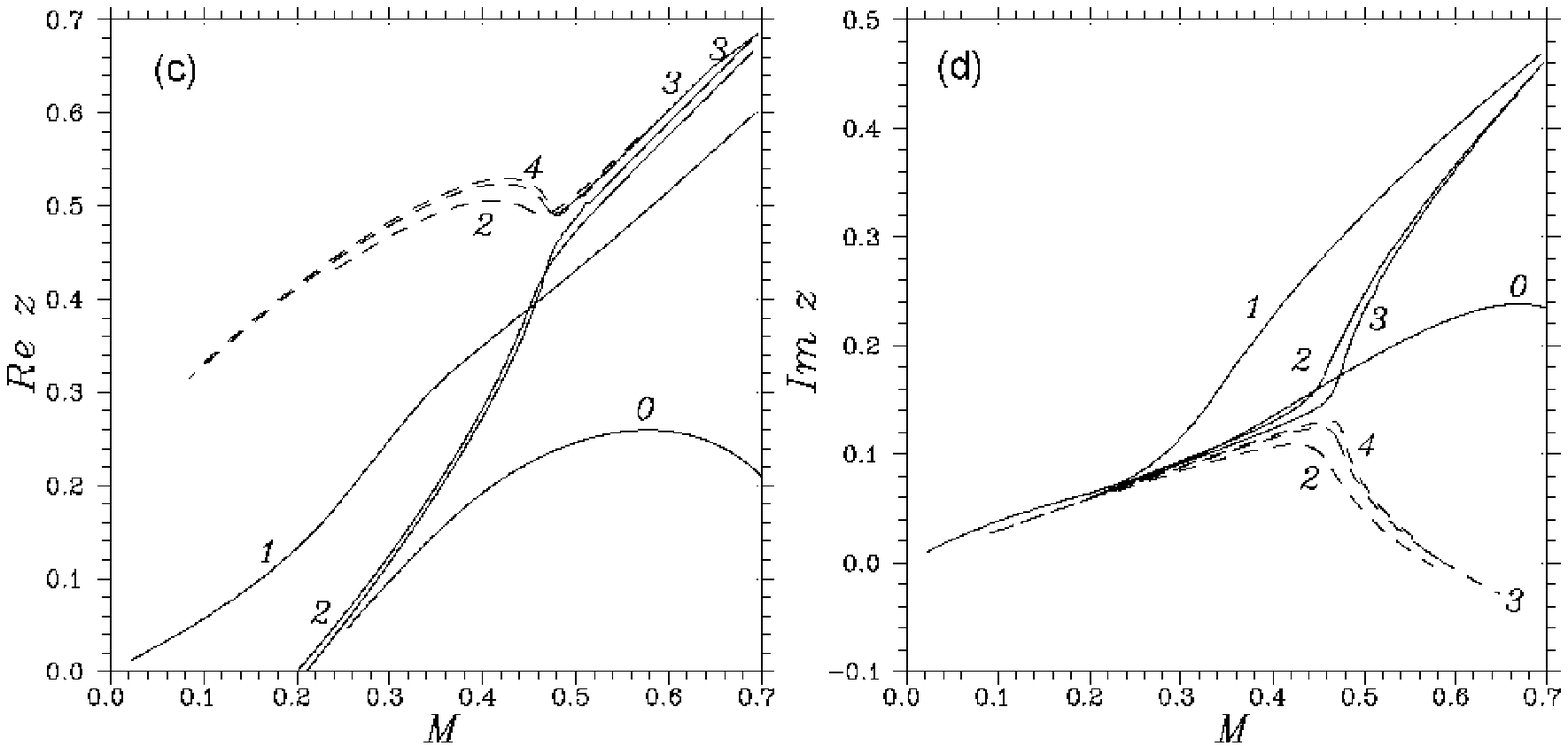}
\caption{ Dimensionless phase velocity $Re\,(\omega / k c_j)$ (a, c) and
amplitude increments  $Im\,(\omega /k c_j)$ (b, d) as functions of the Mach
number $M = V_j / c_j$ for axisymmetric (a, b) and helical $m = 1$ unstable
modes (c, d). The number $n_j$ of zero points of eigenfunctions inside the
jet is indicated near each curve. The jet half-opening angle is equal to
$\theta_j = 20^\circ$, $kr = 5$. The dashed and solid lines show the $u^+$
and $u^-$ mode families, respectively.} \label{f1}
\end{figure*}

\section{Discussion of the results of linear analysis}
\label{4}

\subsection{Medium Without Relaxation}

\begin{figure}[!t]
\includegraphics[scale=1]{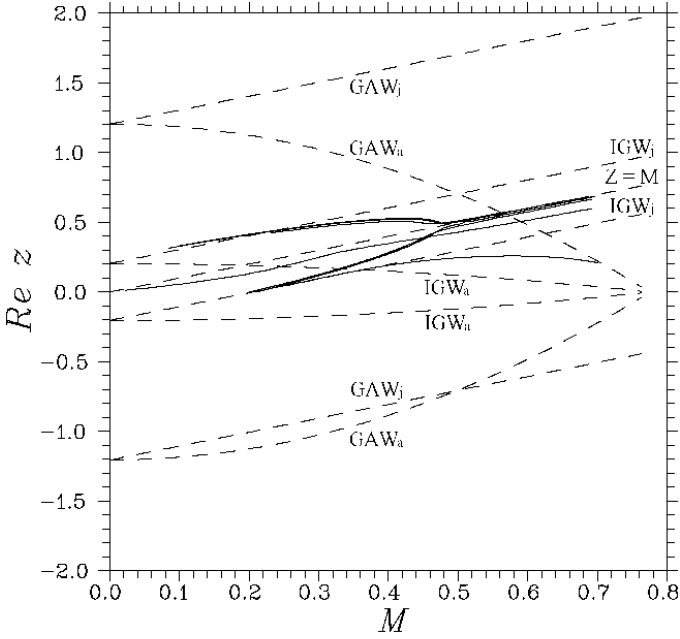}
 \caption{
Dispersion curves from Fig.~\ref{f1}a (solid curves), gravity-acoustic (GAW)
and internal gravity  (IGW) eigenmodes of the jet and atmosphere determined
from eq. (\ref{e37}), (\ref{e38}) (dashed curves).} \label{f2}
\end{figure}

\begin{figure*}[!t]
\includegraphics[scale=1]{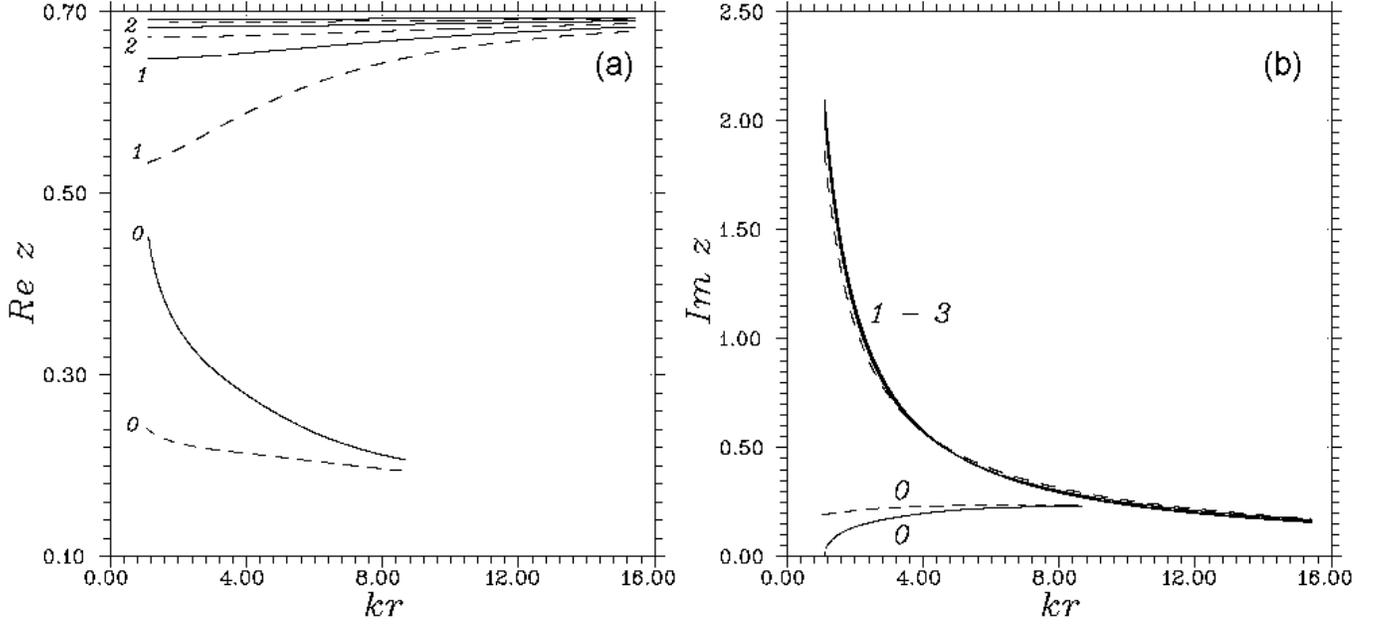}
\caption{Dependences of dimensionless phase velocity $Re\,(\omega / k c_j)$
(a) and fractional growth rate of the amplitude of unstable modes
$Im\,(\omega / k c_j)$ (b) on dimensionless wavenumber $kr$. The solid and
dashed curves show the modes with $m = 0$ and $m = 1$, respectively. The
number of zero points of eigenfunctions inside the jet is indicated near
each curve. The half-opening angle of the jet is $\theta_j = 20^\circ$, $M =
0.7$.} \label{f3}
\end{figure*}

To make it  easier to identify the causes of the development of unstable
modes, we first consider the case of a medium without relaxation: $\tau
\equiv 0$ (adiabatic perturbations). We use the shooting method to solve
numerically the boundary-value problem formulated in Chapter \ref{3}.
Figs.~\ref{f1}--\ref{f3} show the resulting dispersion curves. The spectrum
of unstable modes is discrete and rather complex.

First, the surface modes resulting from the development of the
Kelvin--Helmholtz instability (KHI) in the domain of tangential
velocity discontinuity between the jet and the ambient medium.
These modes decay exponentially with the distance from the jet
boundary on both sides in the $\theta$--coordinate.

Second, the modes of the waveguide made up of the jet boundary,
which are characterized by two ``quantum'' numbers --- $n_j$ and
$m$. Here $n_j$ is the number of nodes of the eigenfunctions of
perturbed pressure between the boundary of the jet and its
symmetry axis determined in the direction perpendicular to this
axis and $m$ is the number of zero points along the jet azimuth
(the number of arms of the helical spiral in the cross section of
the jet). Axisymmetric modes with $m = 0$ are called pinch modes
and nonaxisymmetric modes are called $m$-helical modes according to the
number of zero points in azimuth. The harmonics with no zero
points along the jet radius are the main harmonics and the
remaining ones are reflective harmonics.

Third, the spectrum of modes contains weakly unstable or decaying
( $Im\,\omega<0$) harmonics. For these modes a change of any
parameter has no effect on the number of zero points (i.e., the
$n_a$ of this function in the medium surrounding the jet) of the
$\theta$ distribution of perturbed pressure between the jet
boundary and the $\theta = \pi$ axis, whereas the number $n_j$ of
zero points for the jet does change. Such modes are a direct
consequence of the idealized formulation of the problem, because
the ambient medium can also be formally viewed as a waveguide.
However, in the real situation the development of a standing wave
in  $\theta$ coordinate between the jet boundary and the $\theta
= \pi$ axis shall be impeded by  local heterogeneities present.

In the situation considered the dispersion properties of
perturbations differ substantially from those in the case of the
gravitational potential of a compact object. This is due, first
and foremost, to weak compressibility of the medium in the jet on
the one hand, and to appreciable density stratification on the
other hand.

The local dispersion law allows the existence of two types of
oscillatory modes in each of these media. These are gravity
acoustical waves (GAW) and internal gravity waves (IGW). GAWs are
common longitudinal acoustic waves modified by gradient effects.
IGWs are due to the shear elasticity of the medium resulting from
the disbalance of buoyancy and gravity forces. In the limiting
case of incompressible medium ($M\ll 1$) they become transverse
waves.

Our model has a preferred direction --- that of the velocity shear vector,
which is parallel to the gravity force, --- and therefore for each of these
types of modes one can single out waves propagating in the direction of the
velocity shear and in the opposite direction. Therefore, as is evident from
Fig.~\ref{f1}a,c, all modes break into two families: in the reference frame
connected with matter modes of one family ($u^+$) propagate away from the
ejection source and modes of the other family ($u^-$) propagate toward the
source. In the case of a medium with no radiative cooling all these modes
turn out to be unstable over a wide range of parameters.

One can determine that most of the dispersion curve shown in Fig.~\ref{f1}
corresponds to IGW by directly comparing our results with the solution of
the well-known problem of wave dispersion in a compressible medium with
vertical density stratification due to uniform gravity field $ - g {\bf
e_z}$. The short-wavelength approximation of the corresponding dispersion
equation  has
the following form (see Appendix): %\ref{apa}):

\begin{equation}
\begin{array}{l}
\displaystyle \hat \omega^4 + \bigg[g {d \ln \rho_0 \over d z} -
(k_z^2 + k_\perp^2) c_s^2 \bigg] \hat \omega^2
 \\\displaystyle %\qquad \qquad
- g k_\perp^2 \bigg( g + c_s^2 {d \ln \rho_0 \over d z} \bigg) = 0,
\end{array}
 \label{e35}
\end{equation}
where $\hat \omega = \omega - {\bf kV}$, $k_\perp^2 = k_x^2 +
k_y^2$.

We now use the short-wavelength approximation to make the formal transition
from equation (\ref{e35}) to the case considered by substituting the $r$
coordinate for the $z$ coordinate, $k$ for $k_z$, $k_\perp^2 = k_\theta^2 +
m^2 / r^2$ for $k_\perp^2$, $\rho_i$ for $\rho_0$, $c_i$ for $c_s$, and
$p_i$ for $p_0$. Given the power-law behavior of the dependences of
thermodynamic parameters with exponents (\ref{e8}), we normalize equation
(\ref{e35}) by $k^4 c_i^4$ to find the local dimensionless dispersion laws
for the ambient medium and jet, respectively:

\begin{equation}
\begin{array}{l}
{\displaystyle (z - M)^4 - \bigg( 1 + \delta^2 + {3 \over \gamma
k^2 r^2} \bigg) (z - M)^2  }\\ % \qquad \qquad
\qquad {\displaystyle + {\delta^2 \over \gamma^2 k^2 r^2} (3
\gamma - 1) = 0, }
\end{array}
\label{e37}
\end{equation}

\begin{equation}
\begin{array}{l}
{\displaystyle (\tilde{R}  z)^4 - \bigg( 1 + \delta^2 + {3 \over \gamma k^2
r^2} \bigg) (\tilde{R}  z)^2  }\\ % \qquad \qquad
\qquad {\displaystyle + {\delta^2 \over \gamma^2 k^2 r^2} (3
\gamma - 1) = 0 }.
\end{array}
\label{e38}
\end{equation}
Parameter  $\delta = k_\perp / k$ characterizes the inclination of
the perturbation propagation vector with respect to the radial
direction.

Figure~\ref{f2} shows all solutions of equations (\ref{e37})--(\ref{e38}),
the straight line $z = M$ corresponding to the velocity of the jet matter,
and the dispersion curves from Fig.~\ref{f1}a. It is evident that the
dispersion curves of our jet modes tend to the curves corresponding to IGWs.

Like in the case of jets emerging from young stellar objects
(\cite{ferrari}; \cite{pc}; \cite{hard}; \cite{nh}),  instability in our case
is due to superreflection and superrefraction (\cite{mil}; \cite{rib}).

Flow in the jet is subsonic in our model. However, the excess of velocity
shear at the jet boundary over the wave velocity along this boundary, which
is required for superreflection, is achieved owing to the smallness of the
characteristic propagation velocity of internal gravity waves. For the
parameter values corresponding to the curves shown in
Figs.\ref{f1}--\ref{f3} this corresponds to an $\simeq 0.2$ increase in the
Mach number. Thus the allowance for gravity results in the appearance of
additional unstable jet modes
--- waveguide-resonance internal gravity waves whose
amplification mechanism is due to the superreflection of this type
of waves from the jet boundary.

Because of their low Mach numbers, gravity-acoustic waves do not
satisfy the superreflection condition. In this case the main
physical cause of instability is the Bernoulli effect --- the
well-studied Kelvin--Helmholtz instability develops at the
velocity-shear layer between the jet matter and the surrounding
gas. At the same time, these  waves resonate between the jet
boundary and the $\theta = \pi$ axis and the process is
accompanied by the formation of a weakly unstable wave in the case
of a medium without relaxation.

\begin{figure*}[!t]
\includegraphics[scale=1]{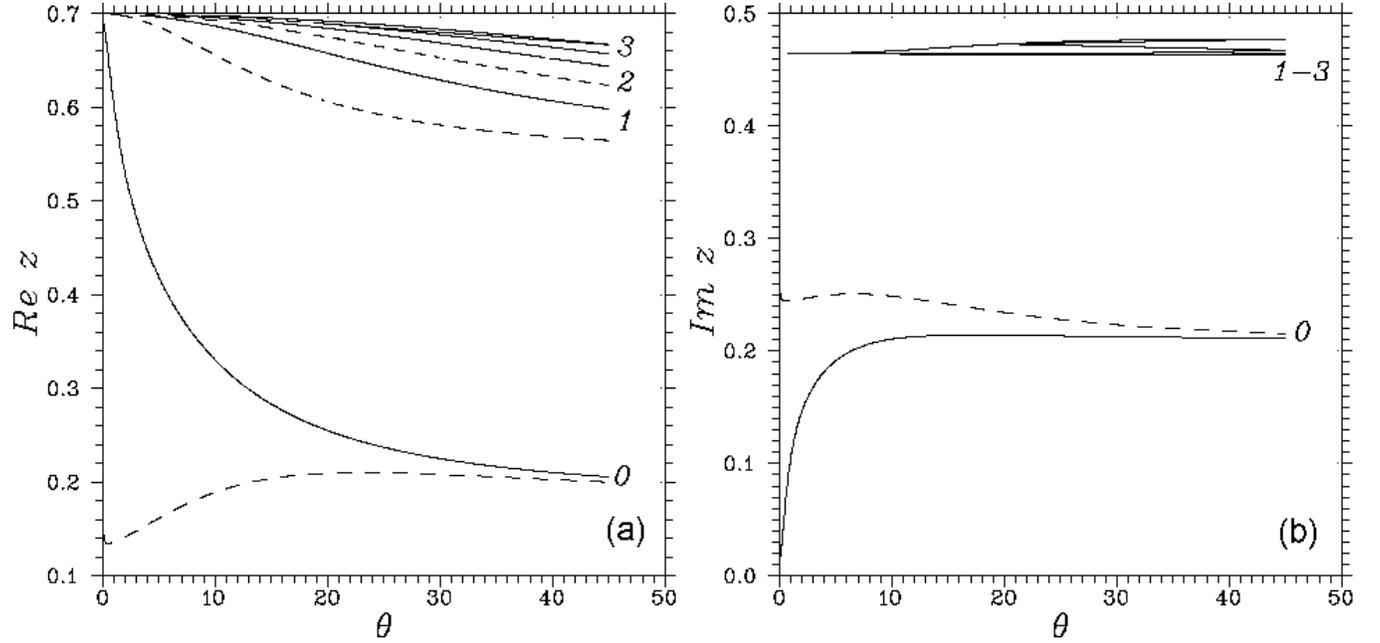}
\caption{ Dimensionless phase velocities $Re\,(\omega / k c_j)$ (a) and
amplitude increments $Im\,(\omega / k c_j)$ (b) as functions of the jet
half-opening angle $\theta_j$  in degrees. Designations are the same as in
Fig. \ref{f3}. $kr = 5$, $M = 0.7$.} \label{f4}
\end{figure*}

The amplitude increment of each reflective harmonic ($n_j \ge 1$) grows
rapidly with decreasing radial wavenumber (see Fig.~\ref{f3}). However, this
is not a physical effect, but only a result of normalization by $k c_j$. It
becomes clear below (see \ref{rad}) that renormalization to the wavenumber
independent V\"{a}is\"{a}l\"{a} frequency results in the unlimited decrease
of the growth time scale of unstable IGW modes with decreasing radial
wavelength.

An important result is that the time scales of the development of the
instability discussed here depend only slightly on the jet opening angle
(see Fig.~\ref{f4}).

The extremely weak dependence of the increments of unstable modes
on jet parameters leads us to suggest that different modes may
develop simultaneously and coexist. On the other hand, we can
conclude that the wavelengths at which the instability builds up,
should be to a greater degree determined by the initial
perturbations.

At the same time, the amplitude increment of reflective harmonics
decreases and that of the main harmonics ($n_j = 0$) --- both the
axisymmetric and the first helical modes of internal gravity waves
--- on the contrary, increases with increasing $kr$. Moreover, at
$kr \ge 10$ these modes are most likely to develop
instability.

\begin{figure*}[!t]
\includegraphics[scale=1]{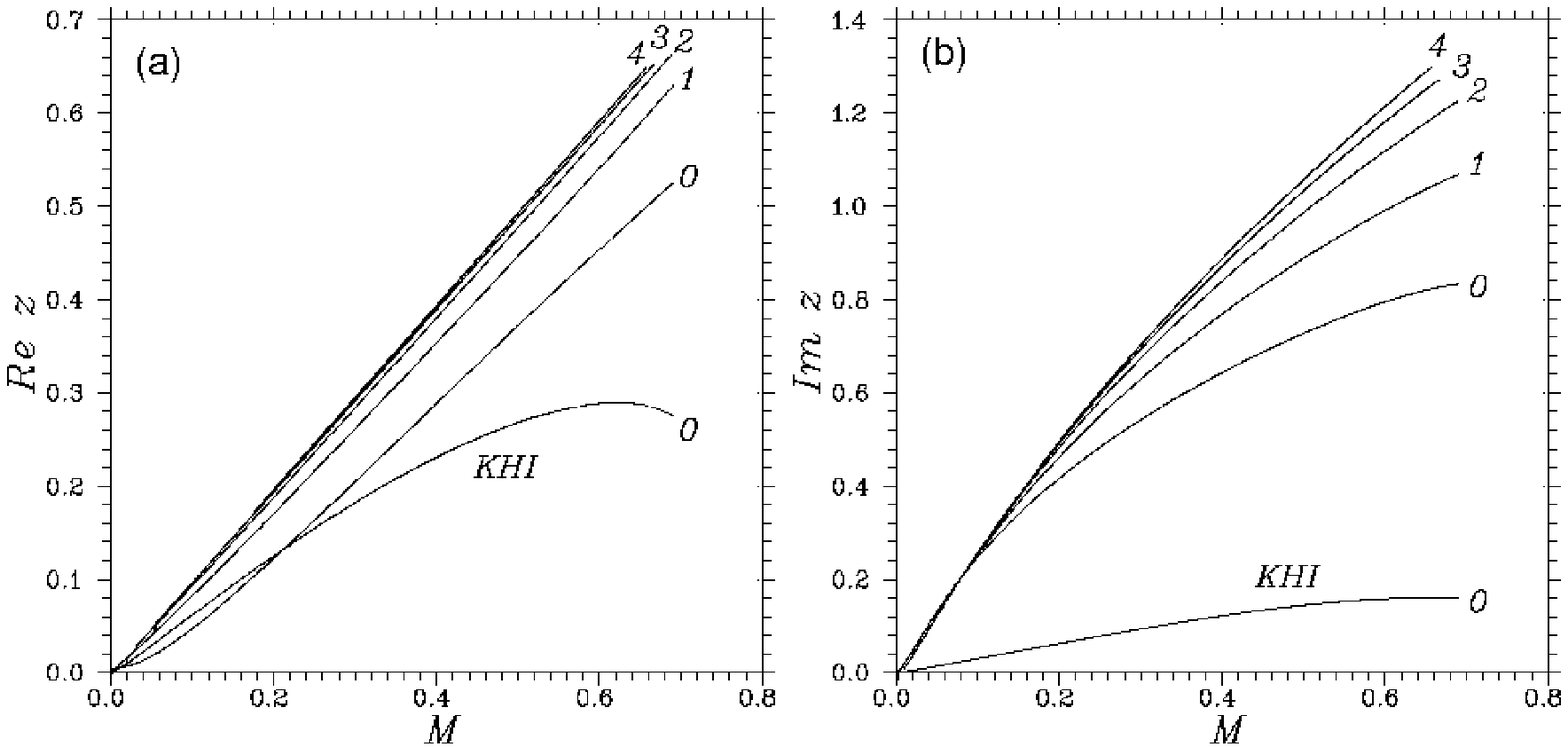}
\caption{Dimensionless phase velocities $Re\,(\omega / k c_j)$ (a) and
amplitude increments $Im\,(\omega / k c_j)$ (b) as functions of the Mach
number for modes with radiative cooling ($\tau = 5$). Each curve is labelled
by the number of the harmonic (the number of zero points of pressure between
the boundary and  the jet axis). Only $u^+$-family harmonics of mode $m =
0$  are shown. $\theta_j = 20^\circ$, $kr = 5$. } \label{f5}
\end{figure*}

\begin{figure*}[!t]
\includegraphics[scale=1]{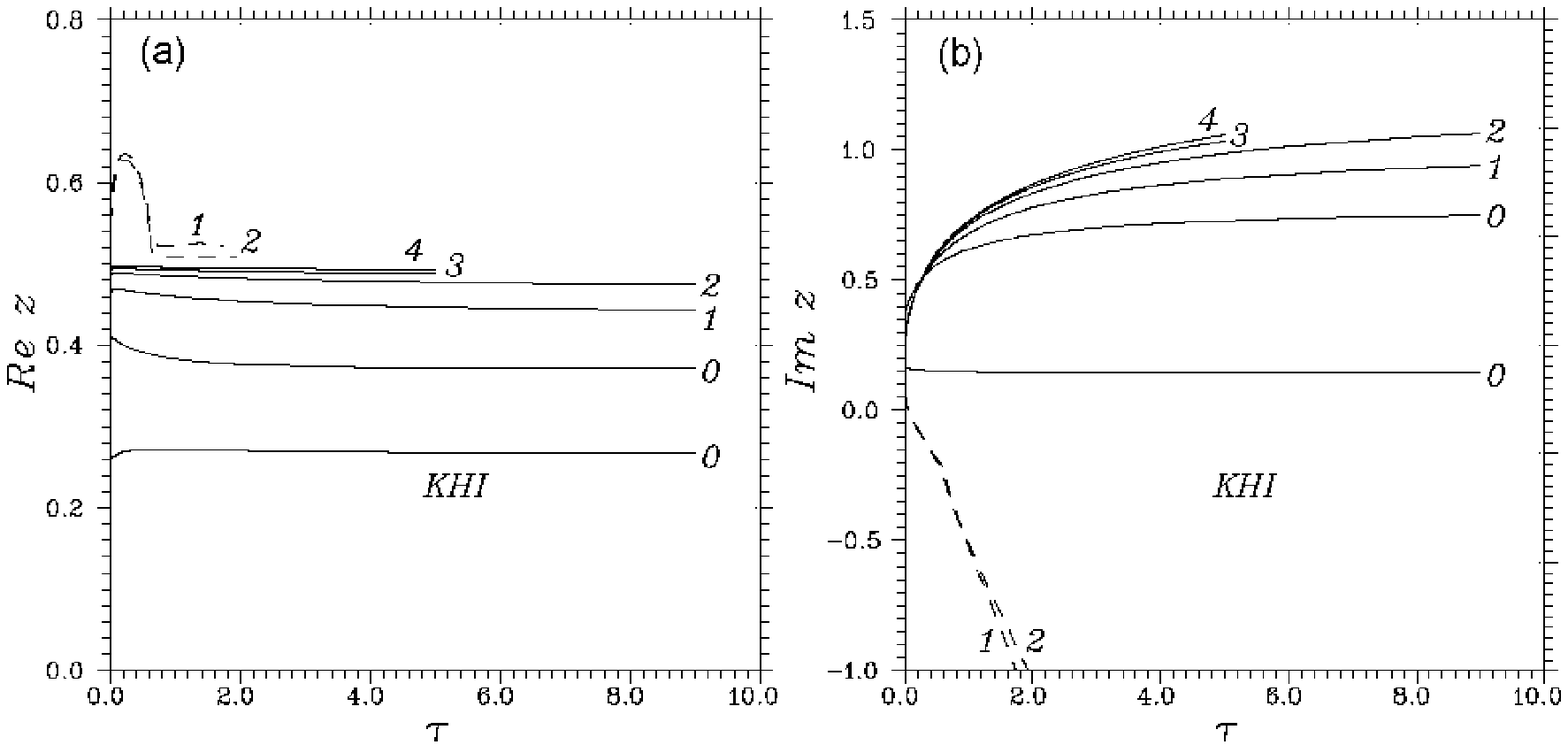}
\caption{ Dimensionless phase velocities $Re\,(\omega /k c_j)$ (a) and
amplitude increments $Im\,(\omega / k c_j)$ (b) as functions of
radiative-cooling parameter $\tau$ for different unstable modes. Each curve
is labelled by the number of the harmonic (the number of zero points of
pressure between the boundary and the jet axis). The dashed lines show the
modes of the $u^+$ family. Only harmonics of mode $m=0$ are shown. $\theta_j
= 20^\circ$, $M = 0.5$, $kr = 5$. } \label{f6}
\end{figure*}

\subsection{Effect of Radiative Cooling on the Dispersion of Unstable Modes}
\label{rad}

The allowance for radiative losses changes radically the spectrum
of unstable modes in the system considered:

{\bf GAW modes} are completely suppressed by radiative cooling and
their decay decrements exceed the frequency by a factor of several
tens to several hundred. This actually means that such modes
cannot even develop. We therefore do not show the corresponding
dispersion curves in our figures in order not to encumber them.

{\bf Surface  KHI modes} are extremely sensitive to radiative cooling. Their
growth time scale increases only slightly compared to the case of adiabatic
perturbations (see Figs. \ref{f5}--\ref{f6}).

\begin{figure*}[!t]
\includegraphics[scale=1]{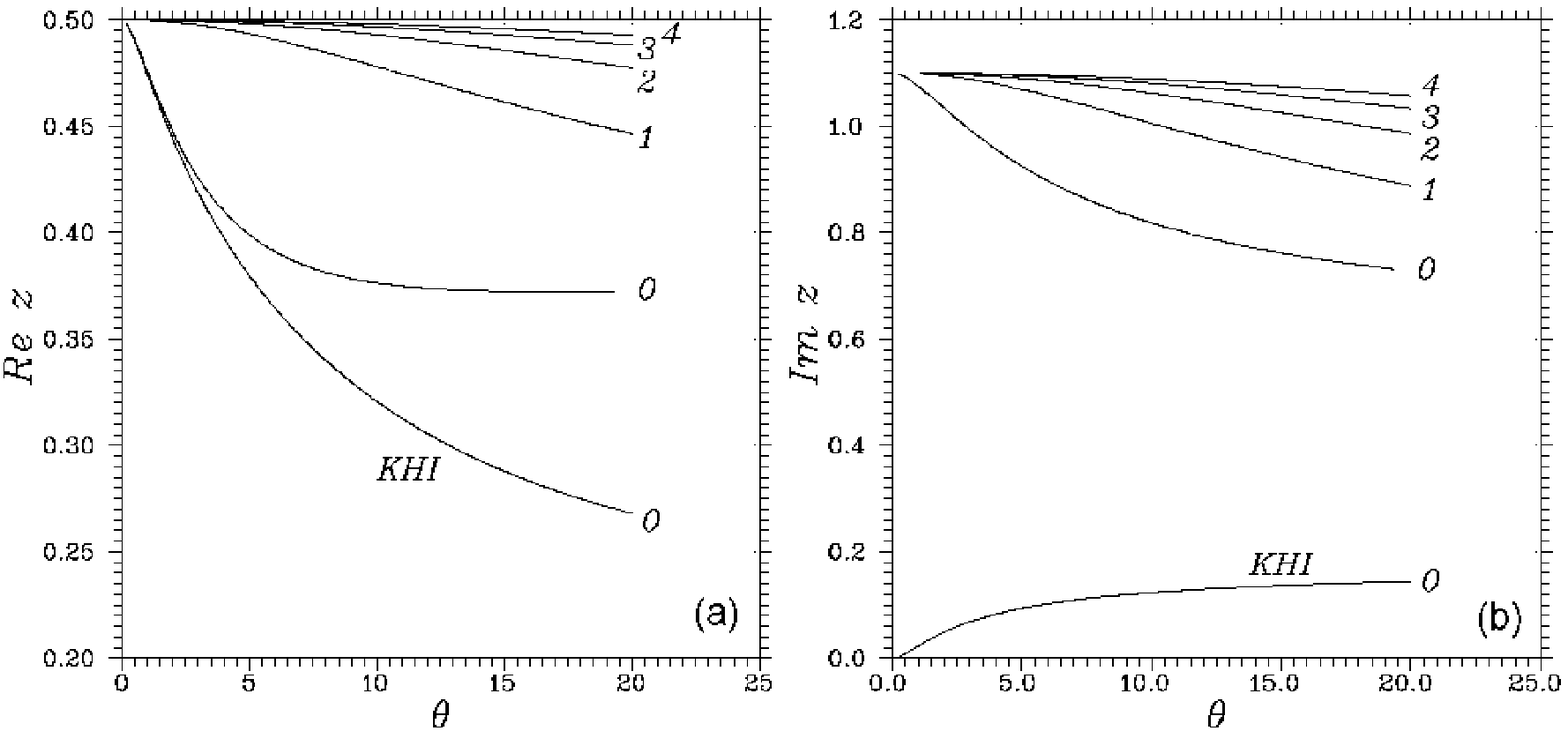}
\caption{Dependences of dimensionless phase velocities $Re\,(\omega / k
c_j)$ (a) and amplitude increments $Im\,(\omega / k c_j)$ (b) on the jet
half-opening angle $\theta_j$ for different unstable modes. Each curve is
labelled by the number of the harmonic (the number of pressure zero points
between the boundary and axis of the jet). Only harmonics of mode $m = 0$
are shown. $kr = 5$, $\tau = 5$.} \label{f7}
\end{figure*}

{\bf Waveguide-resonance modes of IGW of the $\mathbf{u^+}$} family become
damped already at small values of radiation-cooling parameter $\tau$. Their
damping time scale decreases rapidly with increasing $\tau$ (see
Fig.~\ref{f6}b). It can therefore be argued that they are also fully
suppressed by radiative cooling. We show the dispersion curves corresponding
to these modes only in Fig.~\ref{f6}. Note that strictly speaking, in the
presence of radiative cooling these modes should be called
entropy-vortex-type and not internal gravity modes. We nevertheless retain
the old name to facilitate the comparison to the case of the medium without
relaxation.

{\bf Waveguide-resonance IGW modes of the $\mathbf{u^-}$} family increase in
strength substantially because of radiation losses. Their increment
increases with increasing $\tau$ (Fig.~\ref{f6}) without reaching
saturation\footnote{We verified this statement by computations up to $\tau =
100$}.

To explain the latter effect, we must use the formula relating wave energy
density $E$ in a moving medium to wave energy density $E_0$ in the reference
frame comoving with the medium (\cite{ll}):

\begin{equation}
E = E_0 {\omega \over \omega - {\bf k V}}. \label{e39}
\end{equation}
Note that although formula (\ref{e39}) was initially derived for acoustic
waves, it is of universal nature, because it allows simple quantum
interpretation. Namely, the number of photons of the wave field ${\cal N} =
2 \pi E / h \omega = 2 \pi E_0 / h (\omega - {\bf k V})$ does not depend on
the choice of the reference frame (\cite{ll}).

Thus modes of the $u^+$ family, for which $Re\,\omega
> k V_j$ ($Re\,z
> M$), have positive energy density in the jet, whereas modes of the $u^-$ family,
for which $0 < Re\,\omega < k V_j$ ($0 < Re\,z < M$), have
negative energy density. Therefore the decrease of the wave energy
due to radiative cooling decreases the energy of $u^+$-family
modes, and, correspondingly, results in their decay. And vice
versa, the decrease of the wave energy due to radiative cooling
increases the absolute value of the energy density of
$u^-$-family modes and results in their amplification. The latter
situation is a typical example of radiative and dissipative
instability.

Like in the case of a medium without dissipation the dispersion law depends
only very slightly on the change of the jet opening angle over a wide range
of its values (Fig.~\ref{f7}).

\begin{figure*}[!t]
\includegraphics[scale=1]{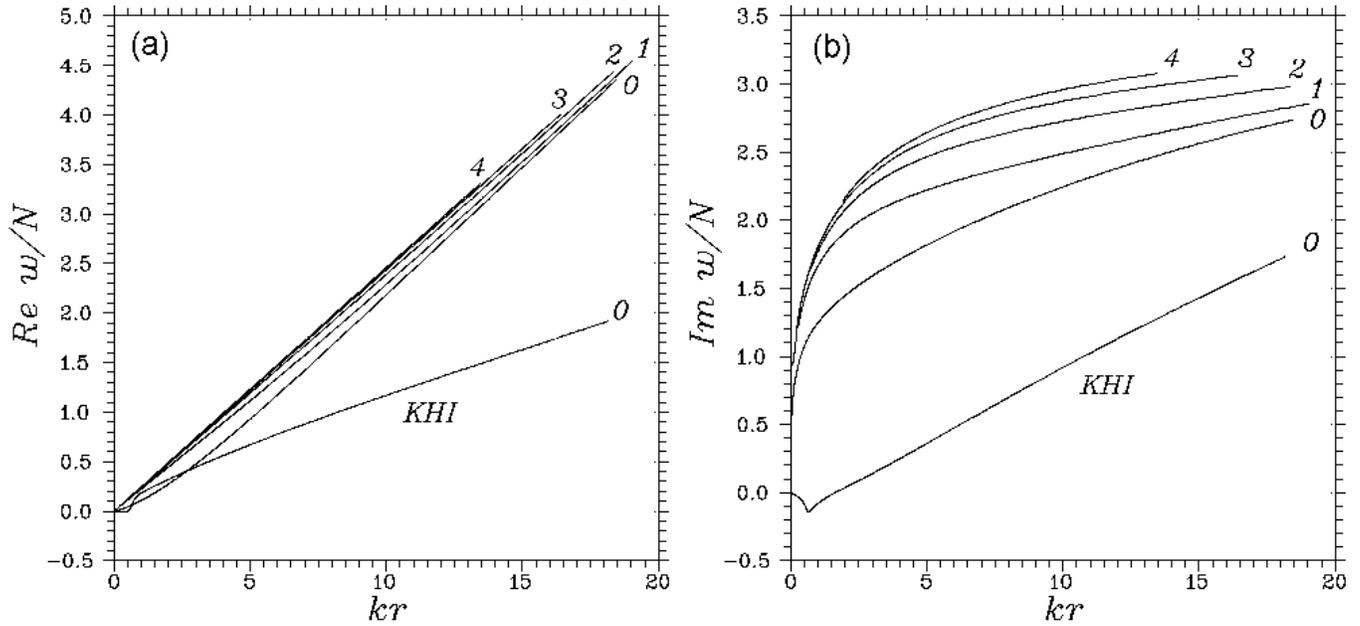}
\caption{ Dependences of dimensionless frequency $Re\,(\omega / N)$ (a) and
increment $Im\,(\omega / N)$ (b) on dimensionless radial wavenumber $kr$ for
different unstable modes. Each curve is labelled by the number of the
harmonic (the number of zero points of pressure between the boundary and
axis of the jet). Only the harmonics of mode $m = 0$ are shown. $M = 0.5$,
$\theta_j = 20^\circ$, $\tau = 5$.} \label{f8}
\end{figure*}

Fig.~\ref{f8} shows how the frequency of unstable perturbations depends on
dimensionless wavenumber $kr$. It is convenient to normalize this frequency
not to the frequency of acoustic waves, which itself depends on $k$, but to
the characteristic frequency of  IGWs --- the Brunt-V\"{a}is\"{a}l\"{a}
frequency:

\begin{equation}
N^2 = {1 \over \rho_j} {d P_j \over d r} \bigg({1 \over \rho_j} {d \rho_j
\over d r} - {1 \over \rho_j c_j^2} {d P_j \over d r} \bigg) = {3 \gamma - 1
\over \gamma^2} {c_j^2 \over r^2}. \label{e40}
\end{equation}

As is evident from Fig.~\ref{f8}b, our analysis predicts unlimited decrease
of the characteristic time scale of the growth of unstable  IGW modes with
decreasing radial wavelength. One must nevertheless bear in mind that the
presence of a transitional layer of finite thickness $l$, where velocity
varies smoothly from $V_j$ inside the jet to zero outside the jet, stabilizes
perturbations with wavelength $\lambda \le l$.

The characteristic feature of the IGW modes considered is that the
maximum of perturbed pressure is achieved at the jet boundary,
whereas both the perturbed displacement $\tilde \xi$ of this
boundary in the direction transversal to it and density
perturbation are equal to zero in the adiabatic case and increase
insignificantly with increasing parameter $\tau$. Moreover,
outside the jet the perturbation amplitudes decrease very rapidly
with the distance from the jet due to the difference of the
impedances of the media: $\rho_j c_j < \rho_a c_a$. We show in
Paper II that during the linear stage of the development of
instability the initial flow (i.e., the jet proper) is not
destroyed because of the smallness of the $\theta$--displacement,
and perturbations remain localized inside a cone near the jet.

In conclusion, we make two small comments. First, although Figs.
\ref{f5}--\ref{f8} show the dispersion curves only for axisymmetric pinch
modes, the results for helical ($m \ge 1$) modes are qualitatively the same.
Second, the computations for bipolar outflows with the subsequent change of
the boundary condition (\ref{e30}) yield the results that are identical to
those that we discuss here for unipolar outflow. This conclusion follows
directly from the localization of unstable modes near the jet boundary that
we discuss in this paper.

\section{Conclusions}

\label{5}

Our linear analysis leads us to conclude that:

\begin{itemize}

\item Conical mass outflows in a field of quadratic gravitational
potential and similar to those observed in a number of Seyfert
galaxies are unstable against the resonance--waveguide development
of a wide spectrum of pinch and helical internal gravity waves.

\item  The characteristic amplitude growth time scale of these
modes depends extremely slightly on the jet opening angle over a
wide range of theses angles.

\item  Radiative cooling suppresses completely all gravitational
acoustic waves, has only a slight effect on unstable surface
Kelvin--Helmholtz modes, and results in the decay of
waveguide--resonance internal gravity modes propagating in
antisource direction relative to the jet matter. And vice versa,
radiative cooling increases substantially the instability of such
sourceward-propagating modes.

\item amplification mentioned above has the form of radiative and
dissipative instability of modes with negative energy density.

\item the formation of the observed regular patterns in radiation
cones in the vicinity of the nuclei of Seyfert galaxies can be due
only to unstable surface modes and slow (sourceward moving in the
jet) waveguide--resonance IGWs modes. The velocities of these
modes along the jet boundary exceed the characteristic sound speed
in the ambient atmosphere and this fact allows us to believe in
their possible evolution into shocks.

\item In the case of small jet opening angles the main harmonic of the pinch mode
of IGWs is most likely to develop in the short-wavelength domain
($kr \ge 20$). In the longer-wavelength domain ($5 \le kr \le
15$) the main harmonic of the first helical mode is most likely
to develop.

\item Because of the different localization of the first helical
and pinch modes the development of one of these modes should no
fatal consequences whatsoever for the other mode.

\end{itemize}

We describe detailed numerical simulations of the development of
these waves in our forthcoming Paper II.

{\appendix
\def\appendixname{Appendix}%
\section{Dispersion relation for  IGWs}

Here we briefly describe the derivation of the dispersion
relation for internal gravity waves. We consider low-amplitude
waves in a compressible medium with vertical density
stratification due to uniform gravitational field $-g{\bf e_z}$,
where $g=const$. The initial linearized set of hydrodynamics
equations has the following form:
\begin{equation}
{\partial\tilde\rho \over \partial t} + \tilde v_z{\partial\rho_0
\over
\partial z}+\rho_0\bigg(\mbox{div}_\bot{\bf v_\bot}+{\partial\tilde v_z \over \partial z}\bigg)
=0, \label{a1}
\end{equation}

\begin{equation}
{\partial\tilde{\bf v}_\bot\over\partial t}=-{1 \over\rho_0}{\bf
\nabla_\bot} \tilde p,
 \label{a2}
\end{equation}

\begin{equation}
{\partial\tilde v_z \over \partial t}=-{1 \over \rho_0}{\partial
\tilde p \over \partial z} -g {\tilde \rho \over \rho_0},
 \label{a3}
\end{equation}

\begin{equation}
{\partial\tilde p \over \partial t} + \tilde v_z{\partial p_0
\over
\partial z}=c_s^2\bigg( {\partial \tilde\rho \over \partial t}+\tilde v_z{\partial \rho_0 \over \partial z}
\bigg). \label{a4}
\end{equation}
Here subscript  $z$ denotes vertical components and subscript
$\bot$, the vector components orthogonal to ${\bf e_z}$. The
equation of hydrostatic equilibrium can be written in the
following form:
\begin{equation}
g=-{1  \over \rho_0(z)} {d p_0(z) \over dz}.
 \label{a5}
\end{equation}
Other designations are the same as in the main part of the paper.

We consider short-wavelength perturbations along the $z$ axis:
\begin{equation}
{1 \over k_z}\bigg\vert{\partial \ln \rho_0 \over \partial z}\bigg \vert
\ll 1,
 \label{a6}
\end{equation}
where $k_z$ is the vertical component of wave vector ${\bf k}$
$({\bf k^2}= k_z^2+{\bf k_\bot^2})$. We then seek a solution in
the form of planar waves:
\begin{equation}
\tilde f(x,y,z,t) = \hat f \exp \{ik_xx+ik_yy+ik_zz-i\omega t\}, \label{a7}
\end{equation}
\noindent where  $\tilde f$ is the perturbed function with amplitude $\hat
f=const$. The set of differential equations (\ref{a1})--(\ref{a4})
transforms into the following algebraic equations:
\begin{equation}
-i\omega\hat\rho+\hat v_z {d\rho_0 \over dz}+ i\rho_0({\bf k_\bot \hat
v_\bot }+k_z\hat v_z)=0, \label{a8}
\end{equation}

\begin{equation}
-i\omega{\bf \hat v_\bot}=-i {\bf k_\bot}{\hat p \over \rho_0},
 \label{a9}
\end{equation}

\begin{equation}
-i\omega  \hat v_z=-i  k_z{\hat p \over \rho_0}-g {\hat \rho \over
\rho_0},
 \label{a10}
\end{equation}

\begin{equation}
-i\omega  \hat p-\rho_0 g \hat v_z =c_s^2\bigg(-i\omega\hat\rho+
\hat v_z{d\rho_0 \over dz}\bigg).
 \label{a11}
 \end{equation}

This equation set is homogeneous and has a nontrivial solution
only if its determinant is identically equal to zero. The latter
condition yields the following dispersion equation:
\begin{equation}
\omega^4+\omega^2\bigg(-k^2c_s^2+g{d\ln\rho_0 \over dz}\bigg)\nonumber \\
-gk_\bot^2\bigg(g+c_s^2{d\ln\rho_0\over dz}\bigg)=0.
 \label{a12}
\end{equation}

In the case of a medium moving with velocity ${\bf V}$ the Doppler transform
for frequency $\hat \omega=\omega-{\bf kV}$ yields equation (\ref{e35}) in
the main part of the paper. }

\begin{acknowledgements}
We are grateful to I.~G.~Kovalenko for his critical comments and to
V.~V.~Levi for numerous useful discussions. The images of the NGC~3516 and
NGC~5252 galaxies were obtained with the 6-meter telescope of the Special
Astrophysical Observatory of the Russian Academy of Sciences funded by the
Ministry of Science of the Russian Federation (registration number 01-43).
This work was supported in part by the ``Nonstationary objects in the
Universe'' program of the Ministry of Industry and Science of the Russian
Federation. A.~V.~Moiseev and V.~L.~Afanasiev also acknowledge the of from
the Russian Foundation for Basic Research (project no. 06-02-16825).
\end{acknowledgements}

%\textit{Translated by V.~Astakhov}

\end{document}